\documentclass[
      ,final
   ]
   {aipproc}

\layoutstyle{6x9}

\usepackage{epsf}

\newif\ifdraft
\drafttrue
\newif\ifpreprint
\preprinttrue

\def\fig#1{fig.~{\ref{#1}}}

\def\eqn#1{eq.~(\ref{#1})}

\def\Fig#1{Figure~{\ref{#1}}}

\def\tr{\, {\rm tr}}
\def\Tr{\, {\rm Tr}}

\def\NeqFour{{\cal N}=4}
\def\Nc{N_c}
\def\Ord{{\cal O}}
\def\e{\epsilon}
\def\del{\partial}
\def\bea{\begin{eqnarray}}
\def\eea{\end{eqnarray}}
\def\be{\begin{equation}}
\def\ee{\end{equation}}
\def\qb{{\overline {\kern-0.7pt q\kern -0.7pt}}}

\begin{document}

\noindent \hfill SLAC--PUB--13167

\title{Gluon scattering in $\NeqFour$ super-Yang-Mills theory from
weak to strong coupling%
\footnote{Talk presented at the 8th International Symposium on 
Radiative Corrections (RADCOR), October 1-5 2007, Florence, Italy.
Research supported by the US Department of Energy under contract
DE--AC02--76SF00515.}
}


\author{Lance J. Dixon
}{address={Stanford Linear Accelerator Center,
        Stanford University,
        Stanford, CA 94309, USA}}

\begin{abstract}
I describe some recent developments in the understanding of
gluon scattering amplitudes in $\NeqFour$ super-Yang-Mills theory in the
large-$\Nc$ limit.  These amplitudes can be computed to high orders in 
the weak coupling expansion, and also now at strong coupling using the 
AdS/CFT correspondence.  They hold the promise of being solvable to all 
orders in the gauge coupling, with the help of techniques based on
integrability.  They are intimately related to expectation values for
polygonal Wilson loops composed of light-like segments.
\end{abstract}

\maketitle

\section{Introduction}

In this talk I would like to describe some remarkable progress that
has been made in the past few years in understanding the structure of 
gauge boson scattering amplitudes in a particular gauge theory, 
$\NeqFour$ super-Yang-Mills theory.  While this theory differs in many
details from the electroweak and QCD theories whose radiative corrections
were the subject of this symposium, there are many common issues,
particularly associated with infrared structure.
Indeed, the understanding of infrared divergences in QCD acquired 
over the last few decades has proved extremely useful in unraveling 
some of the structure of $\NeqFour$ super-Yang-Mills theory.  

$\NeqFour$ super-Yang-Mills theory is the most supersymmetric 
theory possible without gravity.  In the free theory, starting from the 
helicity $+1$ massless gauge boson (``gluon'') state, the four 
supercharges can be used to lower the helicity by $4 \times {1\over2} = 2$ 
units, until the helicity $-1$ gluon state is reached.
If one had more supercharges, one would need spin $>1$ states,
and it is not known how to quantize such theories in a unitary way
without including at least spin 2 gravitons.
Along the way from the helicity $+1$ to the helicity $-1$ gluon state, 
one passes through the 4 massless (Majorana) spin $1/2$ gluinos, 
and 6 real (or 3 complex) massless spin $0$ scalars. 
In this maximally supersymmetric Yang-Mills theory (MSYM),
all the massless states are in the adjoint representation of the gauge
group, which we will take to be $SU(\Nc)$.  The interactions are all
uniquely specified by the choice of gauge group, and one dimensionless
gauge coupling $g$.  The theory is an exactly scale-invariant,
conformal field theory; that is, the beta function vanishes identically 
for all values of the coupling~\cite{MSYMfinite}.

Here we will consider the 't Hooft limit of MSYM,
in which the number of colors $\Nc\to\infty$, with the
't Hooft parameter $\lambda\equiv g^2 \Nc$ held 
fixed~\cite{tHooftlimit}.  In this limit, only planar Feynman 
diagrams contribute.  Also, the anti-de Sitter space / conformal field 
theory (AdS/CFT) duality~\cite{AdSCFT} suggests that
for $\Nc\to\infty$ the weak-coupling perturbation series in $\lambda$
might have some very special properties.   The reason is that,
according to AdS/CFT, the strongly-coupled (large $\lambda$) limit of the 
four-dimensional conformal gauge theory has an equivalent
description in terms of a weakly-coupled string theory.
The intuition is that the perturbative series should know about this simple
strong-coupling limit, and organize itself accordingly~\cite{ABDK}.

\begin{figure}
\includegraphics[width=.8\textwidth, bb = 55 431 526 655 ]{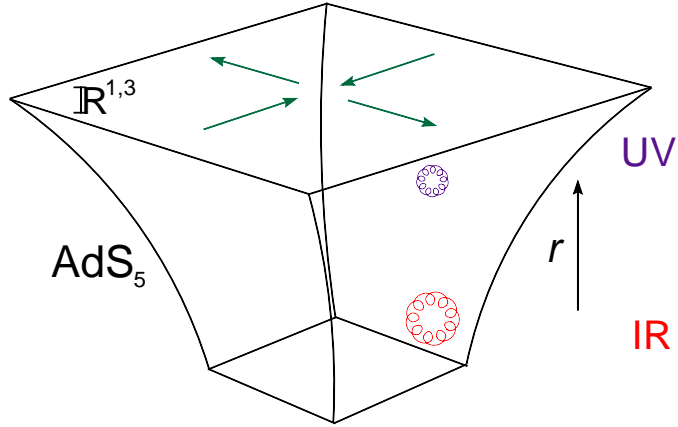}
\caption{Cartoon of the AdS/CFT duality.}
\label{AdS5sketchFigure}
\end{figure}

\Fig{AdS5sketchFigure} sketches how events such as gluon scattering 
look in the AdS/CFT duality~\cite{AdSCFT,AM1}.
Five-dimensional anti-de Sitter space, AdS$_5$, contains,
besides the usual four-dimensional space-time $R^{1,3}$, 
an additional radial variable $r$, which corresponds to a 
resolution scale in the four-dimensional theory.  
Large values of $r$ correspond to the ultraviolet (UV) region; 
small values to the infrared (IR).  The figure shows a ``big''
glueball state in the IR, and a ``small'' glueball state in the UV.
The arrows represent the motion of plane-wave single gluon states
in $R^{1,3}$ for $gg\to gg$ scattering at $90^\circ$.
We'll discuss the motion in $r$ later.  The radius of curvature
of AdS$_5$ is proportional to $\lambda^{1/4}$.  Large $\lambda$
means that the space-time is only weakly curved, which makes it much
simpler to study the string theory; higher excitations of the string can 
usually be neglected.

The AdS/CFT duality is a weak/strong duality.  Quantities that 
can be computed at weak coupling in one picture have a strong-coupling
description in the other picture.  This property makes AdS/CFT both 
powerful and difficult to check explicitly --- although there is 
certainly convincing evidence in its favor.  There are a few quantities 
that are known (modulo a few assumptions) to all orders in $\lambda$; 
that is, for which one can interpolate all the way from weak to strong 
coupling.  Notable among these is the cusp (or soft) anomalous dimension
$\gamma_K(\lambda)$.  The QCD version of this quantity crops up a 
lot in soft-gluon resummation. Beisert, Eden and Staudacher~\cite{BES}
have given an all-orders proposal for $\gamma_K(\lambda)$, based on 
integrability, plus a number of other properties.
Their proposal is consistent with the first four loops in the
weak-coupling expansion~\cite{Neq44,CSV4}, and also
agrees~\cite{BESStrongExpNum,BESStrongExpAnal} with the first three terms
in the strong-coupling 
expansion~\cite{StrongCouplingLeading,StrongCouplingSubleading,RTT}.

In this talk I would like to discuss the evidence for another
proposal~\cite{BDS05}, namely that gluon-gluon scattering $gg\to gg$ 
in MSYM, for any scattering angle $\theta$ can be fully specified 
by just three functions of $\lambda$, independent of $\theta$.
One of these three functions is already ``known'', because it is 
just $\gamma_K(\lambda)$. 
This proposal has received some confirmation at strong coupling,
through the work of Alday and Maldacena~\cite{AM1}.
It was motivated by the structure of IR divergences in gauge theory.

\section{Infrared divergences}

In a conformal field theory, scale invariance implies that the
interactions never shut off, so that a scattering process cannot
really be defined.  While strictly speaking this is true,
we are able to get around it in practice by regulating the theory
in the IR.  We'll use dimensional regularization with $D=4-2\e$ and $\e<0$
(actually a version of it that preserves all the supersymmetry~\cite{FDH}).
The regulator breaks the conformal invariance, but we can recover it by
performing a Laurent expansion around $\e=0$, up to and including the
$\Ord(\e^0)$ terms.

At one loop, there are two types of IR divergences: 
{\it soft}-gluon exchange, in which the virtual gluon energy $\omega \to 0$;
and {\it collinear} regions, in which the gluon's 
transverse momentum (with respect to a massless external line) 
$k_{\rm T} \to 0$.  
The soft and collinear regions each produce a $1/\e$ pole, 
resulting in a $1/\e^2$ leading behavior for on-shell amplitudes at 
one loop.
At $L$ loops, the leading behavior is $1/\e^{2L}$, coming from
multiple soft-gluon exchange that is arranged hierarchically, so
that the outermost gluons are softer and more collinear than the
innermost ones.

In fact, all the pole terms for $L$-loop amplitudes are predictable
in planar gauge theory, thanks to decades of work on the soft/collinear 
factorization and exponentiation of amplitudes, and of 
quark and gluon form factors, in 
QCD~\cite{FormFactors,AmpFact,KM,MagneaSterman,CataniIR,STY}.
For both QCD and MSYM, in the planar limit the pole terms
are given in terms of three quantities (in the notation of 
refs.~\cite{MagneaSterman,STY}):
\begin{itemize}
\item the beta function $\beta(\lambda)$ (but of course this vanishes in
  MSYM),
\item the cusp anomalous dimension $\gamma_K(\lambda)$,
\item a ``collinear'' anomalous dimension $G_0(\lambda)$.
\end{itemize}

The cusp anomalous dimension gets its name because it
appears~\cite{Polyakov,IKR,KR} in the renormalization group 
equation for the expectation value of a Wilson line $W(\rho,g)$
for two semi-infinite straight lines, joined at a kink or cusp:
\be
\biggl( \rho {\del\over\del\rho} + \beta(g) {\del\over \del g}\biggr)
\ln W(\rho,g) = - 2 \gamma_K(\lambda) \, \ln \rho^2 + \Ord(\rho^0),
\label{cuspeq}
\ee
where $\rho^2 \equiv n_1\cdot n_2/(\sqrt{n_1^2 n_2^2}) \to\infty$
as the two straight lines become light-like, $n_1^2, n_2^2 \to 0$.
The cusp anomalous dimension also controls~\cite{KM} the universal 
(flavor independent) large-spin limit of anomalous dimensions 
$\gamma_j$ of leading-twist operators with spin $j$, such as the quark
operators $O_j \equiv \qb (\gamma_+ {\cal D}^+)^j q$:
\be
\gamma_j = {1\over2} \gamma_K(\lambda) \, \ln j + \Ord(j^0)\,, 
\qquad j \to \infty.
\label{gammajcusp}
\ee
Finally, through a Mellin transform of \eqn{gammajcusp},
$\gamma_K(\lambda)$ appears in the large $x$ limit of the DGLAP kernel 
for evolving the parton distributions,
\be
P_{aa}(x) = {1\over2} { \gamma_K(\lambda) \over (1-x)_+ } + \cdots \,,
\qquad x \to 1.
\label{splitcusp}
\ee
Thus, in the study of QCD at colliders it is an important quantity for 
resumming the effects of soft gluon emission.

\begin{figure}
\includegraphics[width=.8\textwidth, bb = -100 1 420 180 ]%
{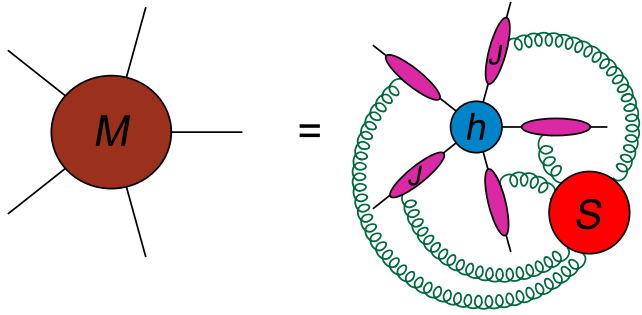}
\caption{Factorization of soft and collinear singularities.}
\label{SoftCollFactFigure}
\end{figure}

The general infrared structure of massless gauge amplitudes
can be exposed~\cite{AmpFact,CataniIR,STY} by factoring 
off soft singularities, which arise
from long-distance gluon exchange, and collinear singularities,
which are also at long distances, but only out along the axis of a
hard parton.  This space-time picture is shown in \fig{SoftCollFactFigure}.
Defining ${\cal M}_n$ to be the full amplitude ${\cal A}_n$ divided
by the tree amplitude ${\cal A}_n^{\rm tree}$,
the factorization formula reads,
\be
{\cal M}_n = S(\{k_i\},\mu,\e) 
\times \prod_{i=1}^n J_i(k_i,\mu,\e) \times h_n(\{k_i\},\mu) \,,
\label{SJh}
\ee
where $\mu$ is the factorization scale, and 
$h_n$ is the hard remainder function, and is finite as $\e\to0$.
The soft function $S$ only sees the classical color charge of the $i^{\rm th}$
particle.  In general it is a complicated matrix acting on the
possible color configurations for $h_n$, because soft gluons can 
attach to any pair of external partons.  The jet function $J_i$ 
is color-diagonal, but depends on the $i^{\rm th}$ spin.
Terms that are color-diagonal and spin-independent can be moved
arbitrarily between $S$ and $J_i$.

\begin{figure}
\includegraphics[width=.8\textwidth, bb = -70 1 500 180 ]%
{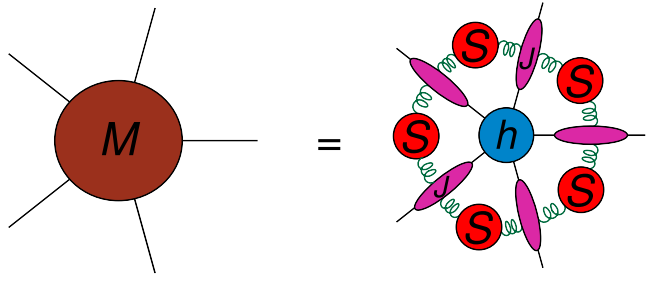}
\caption{Soft-collinear factorization in the planar limit.}
\label{SoftCollFactPlanarFigure}
\end{figure}

In the large-$\Nc$ planar limit, the picture simplifies, to that 
shown in \fig{SoftCollFactPlanarFigure}.  Here $M$ represents
the coefficient of a particular color structure, 
$\tr[ T^{a_1}T^{a_2}\cdots T^{a_n}]$.  
Now soft gluons can only connect adjacent external partons;
and indeed there is no mixing of different color structures at large
$\Nc$.  Because of the color-triviality of the planar limit, one
can absorb the entire soft function $S$ into jet functions, or
break up the right-hand side of \fig{SoftCollFactPlanarFigure} into
$n$ wedges.  Each wedge represents the square root of the Sudakov
form factor, the amplitude ${\cal M}^{[1\to gg]}$ for a color-singlet 
state ``1'' to decay to a pair of partons, say gluons.  
Hence the planar version of \eqn{SJh} is
\be
{\cal M}_n =  \prod_{i=1}^n \biggl[ {\cal M}^{[1\to gg]}
 \biggl({s_{i,i+1} \over \mu^2},\alpha_s,\e\biggr) \biggr]^{1/2}
 \times h_n(\{k_i\},\mu,\alpha_s) \,.
\label{SJhplanar}
\ee
The only dependence of the singular terms on the kinematics is
through the momentum scale, $s_{i,i+1} = (k_i+k_{i+1})^2$,
entering the $i^{\rm th}$ Sudakov form factor.

Factorization also implies that the Sudakov form factor obeys
a differential equation in the momentum
scale~\cite{FormFactors,KR,KM,MagneaSterman},
\be
{\del\over\ln Q^2} \ln {\cal M}^{[1\to gg]}(Q^2/\mu^2,\alpha_s,\e)
= {1\over2} \Bigl[ K(\e,\alpha_s) + G(Q^2/\mu^2,\alpha_s,\e) \Bigr] \,.
\label{SudDiff}
\ee
Here $K(\e,\alpha_s)$ is a pure counterterm, or series of $1/\e$ poles.
By analogy with the $D$-dimensional $\beta$-function,
$\beta(\e,\alpha_s)$, the single poles (related to $\gamma_K$)
determine $K$ completely.
The function $G$ is finite as $\e\to0$, but contains all the $Q^2$
dependence; it will generate a single pole in $\ln {\cal M}^{[1\to gg]}$
upon integrating \eqn{SudDiff} with respect to $Q^2$.
The functions $K$ and $G$ obey renormalization group equations,
\be
\biggl( \mu {\del\over\del\mu} + \beta {\del\over \del g}\biggr) K
= - \biggl( \mu {\del\over\del\mu} + \beta {\del\over \del g}\biggr) G
= - \gamma_K(\lambda).
\label{KGdeq}
\ee
The collinear anomalous dimension $G_0(\lambda)$ arises as a constant
of integration for the differential equation for $G$.

Solving the differential equations for $K$, $G$ and the Sudakov form
factor is particularly easy in a conformal theory because the 
four-dimensional coupling does not run.   Doing this, and inserting the
form-factor solution into \eqn{SJhplanar} for the $n$-point
amplitude, we obtain~\cite{BDS05},
\bea
{\cal M}_n(\e) &=& 1 + \sum_{L=1}^\infty a^L M_n^{(L)}(\e)
\nonumber\\
&=& \exp\Biggl[ - {1\over8} \sum_{l=1}^\infty a^l 
\biggl( { \hat{\gamma}_K^{(l)}\over (l\,\e)^2 }
      + {2 \, \hat{G}_0^{(l)} \over l\,\e } \biggr)
  \sum_{i=1}^n \biggl({\mu^2\over-s_{i,i+1}}\biggr)^{l\,\e}
 \Biggr] \times h_n(\{k_i\}) \,,
\label{Mplanarsing}
\eea
where 
\be
a \equiv { \Nc \alpha_s \over 2\pi} (4\pi e^{-\gamma})^\e
   = {\lambda\over8\pi^2} (4\pi e^{-\gamma})^\e
\label{adef}
\ee
is the loop expansion parameter in the 't Hooft limit,
and $\hat{\gamma}_K^{(l)}$ and $\hat{G}_0^{(l)}$ are the $l$-loop
coefficients of $\gamma_K(a)$ and $G_0(a)$.

The argument of the exponential in \eqn{Mplanarsing}
looks very much like the one-loop amplitude, but with $\e$
replaced by $l\,\e$, denoted by $M_n^{(1)}(l\e)$.  
Thus we are motivated to rewrite \eqn{Mplanarsing} as
\be
{\cal M}_n(\e) = \exp\biggl[ \sum_{l=1}^\infty a^l \Bigl(
  f^{(l)}(\e) M_n^{(1)}(l\e) + h_n(\{k_i\}) + \Ord(\e) \Bigr) \biggr] \,,
\label{Mplanarnew}
\ee
where $f^{(l)}(\e) \equiv f_0^{(l)} + \e \, f_1^{(l)} + \e^2 \, f_2^{(l)}$
collects three series of constants.  Two of these are identified with
the previous quantities as,
\be
f_0^{(l)} = {1\over4} \hat{\gamma}_K^{(l)}, 
\qquad
f_1^{(l)} = {l\over2} \hat{G}_0^{(l)}, 
\label{f01def}
\ee
while the third quantity, $f_2^{(l)}$, is related to the consistency 
of \eqn{Mplanarnew} under collinear limits~\cite{ABDK}.

\section{A surprising relation}

The surprise in planar MSYM is that in some cases the hard remainder 
function $h_n(\{k_i\})$ defined through \eqn{Mplanarnew} 
is actually a constant, independent of the kinematics.  
This result, which has been tested perturbatively for $n=4$ through three 
loops~\cite{ABDK,BDS05}, and for $n=5$ at two loops~\cite{TwoLoopFive},
is a conjecture beyond that:
\be
{\cal M}_n = \exp\biggl[ \sum_{l=1}^\infty a^l \Bigl(
  f^{(l)}(\e) M_n^{(1)}(l\e) + C^{(l)} + \Ord(\e) \Bigr) \biggr] \,.
\label{BDSansatz}
\ee
The dependence of the finite part of the logarithm of the amplitude
is predicted to all orders by \eqn{BDSansatz}, in terms 
of the cusp anomalous dimension.  The prediction for four-gluon
scattering is 
\be
{\cal M}_4^{\rm finite} = \exp\biggl[ {1\over8} \gamma_K(a)
  \ln^2\biggl({s\over t}\biggr) + {\rm const.}\biggr] \,,
\label{BDSfiniteterms}
\ee
where $s=s_{12}$, $t=s_{23}$.
As we shall discuss in section~\ref{strongsection}, 
this formula was confirmed at strong coupling by Alday and Maldacena~\cite{AM1}
using the AdS/CFT correspondence~\cite{AdSCFT}.
In contrast, even at two loops there does not appear to be any 
comparably simple formula for the finite parts of four-gluon 
scattering amplitudes in QCD, or for the subleading-in-$\Nc$ 
terms in MSYM~\cite{ABDK}.
Instead of a constant, as in \eqn{BDSansatz}, one finds that $h_n^{(2)}$
in \eqn{Mplanarnew} is given by a complicated combination of 
polylogarithms involving the dimensionless ratio $t/s$.  On the other
hand, \eqn{BDSansatz} is reminiscent of the
observation~\cite{EynckLaenenMagnea} that finite terms can also 
exponentiate in QCD, in {\it e.g.} the Drell-Yan cross section near
partonic threshold.

\begin{figure}
\includegraphics[width=.8\textwidth, bb = -40 1 460 180 ]%
{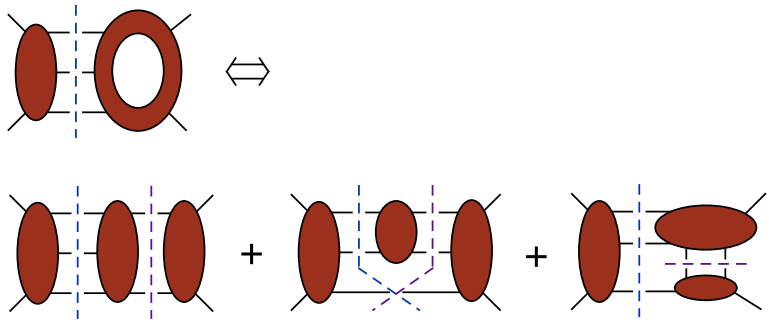}
\caption{Example of generalized unitarity at three loops.}
\label{GenUnitarityFigure}
\end{figure}

\section{Evidence}

The evidence in favor of \eqn{BDSansatz} was collected
from explicit computations of the multi-loop scattering amplitudes.
The amplitudes were constructed by evaluating
(generalized) unitarity cuts~\cite{ELOP,Neq4Oneloop,Neq1Oneloop,Zqqgg,%
GenUnitarityMultiloop,BCFGeneralized}
and matching them to compact representations in terms of a relatively
small number of multi-loop integrals, which turn out to have
rather interesting properties.
Ordinary unitarity relates discontinuities (cuts) in a given channel 
to products of lower-loop amplitudes, summed over
the possible intermediate states in that channel.  
Generalized unitarity allows the lower-loop amplitudes to be 
further sliced, all the way down to tree amplitudes.
\Fig{GenUnitarityFigure} shows an ordinary three-particle cut
for the four-gluon amplitude.  The information in this cut
can be extracted more easily by further cutting the one-loop 
five-point amplitude on the right-hand side of the cut, 
decomposing it into the product of a four-point tree and a 
five-point tree; as illustrated, there are three inequivalent ways to do this.
If one finds a representation of the amplitude that reproduces
all the generalized cuts (in $D$ dimensions), then that representation
is correct.

\begin{figure}
\includegraphics[width=.95\textwidth, bb = 20 460 590 710 ]%
{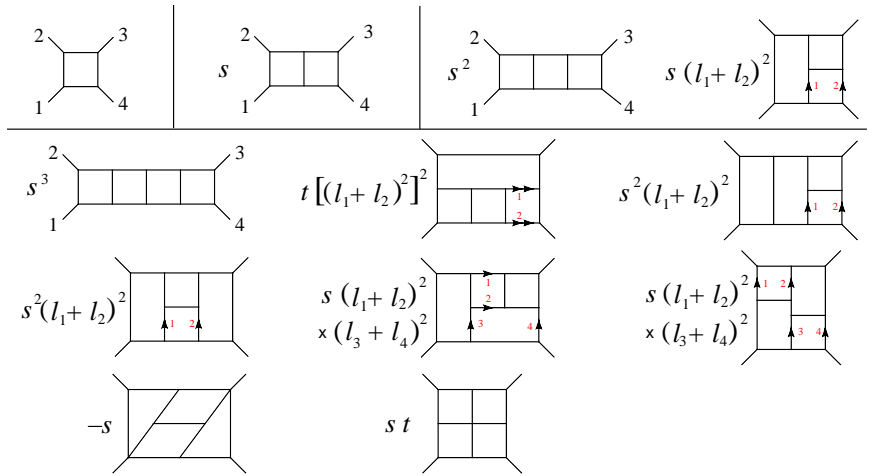}
\caption{Integrals contributing to four-gluon scattering in planar MSYM,
from one to four loops.}
\label{OneToFourFigure}
\end{figure}

\Fig{OneToFourFigure} shows the integrals that enter the four-gluon
scattering amplitude in planar MSYM, from one to four
loops~\cite{BRY,Neq44}, along with their numerator factors.
An overall factor of $st$ is omitted from the rescaled amplitude 
${\cal M}_4(s,t)$, and only one permutation of each integral is shown.
At one and two loops, only scalar integrals appear; that is, the numerator
factors in the integrand depend only on the external momentum invariants.
At three loops, there are two integrals, the scalar triple ladder integral
and the ``tennis-court'' integral shown at the top right of
\fig{OneToFourFigure}.  The latter integral marks the first
appearance of a loop-momentum factor in the numerator, of the form
$(l_i+l_j)^2$, as dictated by the ``rung rule''~\cite{BRY}.
The rung-rule correctly describes all integral topologies that can 
be reduced to trees by a sequence of two-particle cuts.  At four loops,
the last two integrals in \fig{OneToFourFigure} have no two-particle cuts,
and are somewhat more work to determine.  At five loops (not shown)
there are a total of 34 distinct integrals~\cite{FiveLoop}.
Still, it is remarkable that so few integrals are required to describe
the amplitude.

\begin{figure}
\includegraphics[width=.45\textwidth, bb = 35 610 235 720 ]%
{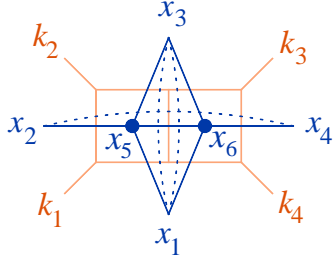}
\caption{The two-loop planar double box integral (in orange) and
associated dual graph (in blue).}
\label{TwoLoopDualFigure}
\end{figure}

\section{Pseudo-conformal integrals}
\label{pseudoconfsection}

In fact, the integrals that appear in the four-point amplitude through
five loops are all {\it pseudo-conformal}.  To describe what this 
means~\cite{MagicIdentities},
first consider taking all the external legs off shell, $k_i^2 \neq 0$,
in order to be able to perform the integral without dimensional
regularization, in $D=4$.  Next define dual momentum or sector variables $x_i$,
such that the original momentum variables $k_i$ are differences
of the $x_i$, with $k_i^\mu = x_{i+1}^\mu - x_i^\mu$.  
Similarly define an $x_i$
associated with each loop, such that $x_{ij} \equiv x_i - x_j$
is equal to the momentum flowing through the propagator that separates
$x_i$ from $x_j$.  \Fig{TwoLoopDualFigure} illustrates the dual diagram (in blue)
associated with the planar double box integral (in orange) which appears
in the two-loop MSYM amplitude.  The dual propagators (denominator factors)
are shown as solid blue lines, while dashed blue lines correspond to
numerator factors in the integrand.  The integral is given by
\bea
I^{(2)}(\{k_i\}) &=& s^2 t \int { d^4p\ d^4q 
\over p^2 (p-k_1)^2 (p-k_1-k_2)^2  q^2 (q-k_4)^2 (q-k_3-k_4)^2 (p+q)^2 }
\label{DoubleBox}\\
&=& (x_{13}^2)^2 x_{24}^2 \int { d^4 x_5\ d^4 x_6 
\over x_{15}^2 x_{25}^2 x_{35}^2  x_{36}^2 x_{46}^2 x_{16}^2 x_{56}^2} \,,
\label{DoubleBoxDual}
\eea
using $s = (k_1+k_2)^2 = x_{13}^2$, $p^2 = x_{15}^2$, and so forth.

Under an inversion, $x_i^\mu \to x_i^\mu/x_i^2$, we have
\be
x_{ij}^2 \rightarrow {x_{ij}^2 \over x_i^2 x_j^2} \,, \hskip 1 cm
 {d^4 x_5} \rightarrow {d^4 x_5 \over (x_5^2)^4} \,, \hskip 1 cm
{d^4 x_6} \rightarrow {d^4 x_6 \over (x_6^2)^4} \,,
\label{inversionsym}
\ee
and it is easy to see that \eqn{DoubleBoxDual} is left invariant.
In general, an integral is invariant under inversion if there is a net 
of zero (four) lines emerging from each external (internal) $x_i$ vertex,
where ``net'' means solid lines minus dashed lines.
Every integral is automatically invariant under translations of the
dual variables, $x_i \to x_i + c$, and under Lorentz transformations.  
Because these transformations, together with inversions, generate the
conformal group, invariance under inversion suffices to guarantee 
dual conformal invariance for the integral.
Now we can define a pseudo-conformal integral to be one which
is finite in $D=4$, after all the $k_i^2$ are taken off-shell,
is dual conformal invariant, and possesses a smooth $k_i^2\to0$ limit.
The last condition ensures that the integral does not become infinite
or vanish as we return to the on-shell limit.

Dual conformal symmetry arose in the context of multi-loop ladder 
integrals~\cite{Broadhurst}, and in two dimensions in the 
theory of (planar) Reggeon interactions~\cite{LipatovDuality}.
Its relevance for the structure of MSYM amplitudes was first pointed out
by Drummond, Henn, Smirnov and Sokatchev~\cite{MagicIdentities},
based on the structure of the amplitudes through three loops, 
and the rung-rule contributions at four loops.  
The four- and five-loop four-gluon amplitudes can be organized as well, 
according to the two principles:
\begin{itemize}
\item Only pseudo-conformal integrals appear.
\item The pseudo-conformal integrals appear only with weight $\pm1$.
\end{itemize}
Originally it appeared that two integrals at four loops~\cite{Neq44} 
and 25 integrals at five loops~\cite{FiveLoop} were pseudo-conformal 
but did {\it not} appear in the amplitude.  However, it was later pointed out that 
those integrals were not actually finite in $D=4$~\cite{DrummondVanishing}.
Recently, some intuition into the signs $\pm1$ has been given by 
considering the singularity structure of the various integrals more
carefully~\cite{CachazoSkinner}.

\section{Evaluating integrals}

Once the structure of the amplitude is known in terms of basic integrals,
the next task is to evaluate those integrals, analytically if possible,
otherwise numerically.  For example, to test \eqn{BDSansatz} at 
three loops, we first expand it out to third order, obtaining
the iterative relation,
\be
M_n^{(3)}(\e) = - {1\over3} \Bigl[ M_n^{(1)}(\e) \Bigr]^3 
+ M_n^{(1)}(\e) M_n^{(2)}(\e) + f^{(3)}(\e) M_n^{(1)}(3\e)
 + C^{(3)} + \Ord(\e).
\label{BDS3}
\ee
To test this relation at order $\e^0$ for $n=4$~\cite{BDS05}, 
we need the following integrals:
\begin{itemize}
\item The one-loop box integral through $\e^4$ --- because
it has $1/\e^2$ poles, and appears cubed in \eqn{BDS3}.
\item The planar double box integral~\cite{SmirnovDoubleBox} 
in \fig{TwoLoopDualFigure} through $\e^2$ --- because $M_4^{(2)}(\e)$ 
appears in \eqn{BDS3} multiplied by $M_4^{(1)}(\e)$.
\item The triple ladder~\cite{SmirnovTripleBox}
and tennis-court~\cite{BDS05} integrals through $\e^0$.
\end{itemize}
Mellin-Barnes techniques (see {\it e.g.} ref.~\cite{SmirnovBook})
are very useful in this regard.
Inserting the results into \eqn{BDS3}, and using identities
among weight 6 harmonic polylogarithms~\cite{HPL}, the relation~(\ref{BDS3})
was verified, and three of the four constants at three loops
could be extracted:
\bea
f_0^{(3)} &=& {11\over5} (\zeta_2)^2,
\qquad
f_1^{(3)} = 6 \zeta_5 + 5 \zeta_2 \zeta_3,
\qquad
f_2^{(3)} = c_1 \, \zeta_6 + c_2\, (\zeta_3)^2,
\label{f3loop}\\
C^{(3)} &=& \biggl( {341\over216} + {2\over9} c_1 \biggr) \zeta_6
  + \biggl( -{17\over9} + {2\over9} c_2 \biggr) (\zeta_3)^2 \,.
\label{C3loop}
\eea
The first two of these constants control infrared divergences.
The value of $f_0^{(3)} = \hat{\gamma}_K^{(3)}/4$ 
confirms a result for the three-loop cusp
anomalous dimension in planar MSYM, which was first obtained~\cite{KLOV} 
by applying the principle of ``maximal transcendentality'' to the 
corresponding result in QCD~\cite{MVV}.
The value of $f_1^{(3)} = (3/2) \hat{G}_0^{(3)}$ 
gives the three-loop collinear anomalous
dimension, which was found to agree (applying the same principle)
with the QCD result~\cite{ThreeLoopQCDFormFactor}.
The constants $f_2^{(3)}$ and $C^{(3)}$ are inseparable using only
the four-gluon amplitude; either the five-gluon amplitude or
a collinear analysis would be required to separate them.
The numbers $c_1$ and $c_2$ are expected to be rational.

A similar analysis can be performed at four 
loops~\cite{Neq44,CSV4,CSVCollinear}, except that the integrals
become less tractable analytically.
Fortunately, there are methods available for automating the construction
of Mellin-Barnes representations~\cite{AMBRE}, the extraction of
$1/\e$ poles, and the setting up of numerical integration over multiple
contours for the Mellin inversion~\cite{AnastasiouDaleo,CzakonMB}.
Before describing the four-loop results, let us turn to some
very interesting developments that have taken place, based 
on integrability.

\section{Integrability and anomalous dimensions}

\begin{figure}
\includegraphics[width=.8\textwidth, bb = 60 560 531 720 ]{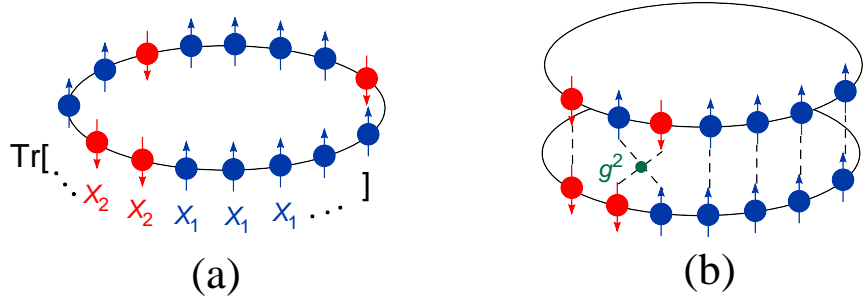}
\caption{(a) Mapping a single-trace operator to a spin chain.
(b) One-loop contribution to the anomalous dimension matrix at large
  $\Nc$.}
\label{SpinChainFigure}
\end{figure}

In large-$\Nc$ gauge theory, a preferred role is played by local 
``single-trace operators''.  In the case of MSYM, one subsector
of such operators is provided by products of the 3 complex scalar fields,
$X_i$, $i=1,2,3$.  The operator $\Tr[ X_1^n ]$ is a so-called BPS
operator, and is unrenormalized to all orders in $\lambda$.  
A set of operators with more interesting renormalization properties 
are close to BPS~\cite{BMN}, 
and contain $X_2$ fields as well as $X_1$, for example,
\be
\Tr [ \ldots X_2 X_2 X_1 X_1 X_1 \ldots ] \,.
\label{singletracesu2}
\ee
As shown in \fig{SpinChainFigure}(a), this set of operators can be mapped
to a one-dimensional, periodic spin chain, in which $X_1$ ($X_2$) is
mapped to spin up (spin down), corresponding to a finite-dimensional 
(spin 1/2) representation of $SU(2)$ spin symmetry.

The anomalous dimensions of the set of operators~(\ref{singletracesu2})
are found by diagonalizing the dilatation operator, which can be
mapped to a Hamiltonian for the spin chain.  In the large-$\Nc$ limit,
this Hamiltonian is local, because non-local interactions correspond 
to non-planar diagrams.  For example, as shown in
\fig{SpinChainFigure}(b), a one-loop contribution from a four-scalar
interaction can only affect color-adjacent $X_i$ fields (spins).
(The range of the interactions does increase with the number of loops.)
Minahan and Zarembo~\cite{MinahanZarembo} showed that the one-loop
Hamiltonian was {\it integrable}; that is, the system possesses
\begin{itemize}
\item infinitely many conserved charges,
\item a spectrum of quasi-particles (spin waves, or magnons),
\item magnon scattering via a $2\to2$ $S$ matrix obeying the
  Yang-Baxter equation,
\item solutions for the anomalous dimensions (energies) via a Bethe ansatz.
\end{itemize}
Integrable structures in QCD had been identified 
previously~\cite{QCDHEIntegrable,QCDIntegrable,BGK}.  In planar MSYM, 
however, the integrability appears to persist to all orders in $\lambda$; 
indeed, it is known to be present at strong coupling, 
from the form of the classical sigma model on target space 
AdS${}_5\times S^5$~\cite{BPR}.

There is a rich literature of extensions of the one-loop results
of ref.~\cite{MinahanZarembo} to higher loops, even all loop orders, 
and to more general sectors of planar MSYM, which I can only touch on
here~\cite{BES,BKSS,BS05,B05,EdenStaudacher}.
The sector most relevant to gluon scattering amplitudes is not the
spin 1/2 $SU(2)$ sector~(\ref{singletracesu2}), but that in which 
the $X_2$ fields are replaced by covariant derivatives ${\cal D}^+$ 
acting in the $+$ (light-cone) direction,
\be
\Tr [ \ldots {\cal D}^+ {\cal D}^+ X_1 X_1 X_1 \ldots ] \,.
\label{singletracesl2}
\ee
These derivatives act as an infinite-dimensional representation
of the noncompact version of $SU(2)$, namely $SL(2)$.
Within this sector, the cusp anomalous dimension can be found
by taking the limit of a small number of fields (spin chain length) $L$,
and a large number of derivatives $j$, to get the operator
\be
O_j = \Tr\bigl[ X_1 ({\cal D}^+)^j X_1 \bigr], \quad j\to\infty.
\label{cuspscalar}
\ee
By the universality of the cusp anomalous dimension, it does not matter
which leading-twist large $j$ operator is used; they all have the 
behavior~(\ref{gammajcusp}) at large $j$.

\section{An all-orders proposal}

In brief, and omitting many subtleties,
the Bethe-ansatz solution consists of taking the eigenstates
of the Hamiltonian to be multi-magnon states, with phase-shifts induced
by repeated $2\to2$ scatterings.  The periodicity of the wave
function on the closed chain leads to the Bethe condition, which depends
on the chain length $L$.  In the limit $L\to\infty$, the Bethe condition
becomes an integral equation, which depends on the form of the 
$2\to2$ magnon $S$ matrix~\cite{EdenStaudacher}.  
This $S$ matrix is {\it almost} fixed by the symmetries, but an 
overall phase, the {\it dressing factor}, is not so easily deduced.  
Finally, there is a potential {\it wrapping problem} in extrapolating
to the cusp anomalous dimension:  The Bethe ansatz is only rigorously
valid when the interaction range (the number of loops) is smaller than
the chain periodicity $L$.  However, even though the cusp anomalous dimension
has $L=2$, it has been argued that its universality leads it to 
appear within large-$L$ sectors, and renders it immune to the wrapping 
problem~\cite{BGK,EdenStaudacher,DressWrap}.

Eden and Staudacher~\cite{EdenStaudacher} derived an integral equation
for the all-orders behavior of the cusp anomalous dimension
from an all-loop Bethe ansatz~\cite{BS05},
by assuming that the dressing factor did not play a role perturbatively.
This equation agreed with the known one-, two-, and three-loop coefficients
of $\gamma_K(\lambda)$, and made the four-loop prediction,
\be
f_0^{(4)} \Big|_{\rm ES} = 
{1\over4} \hat{\gamma}_K^{(4)} \Big|_{\rm ES}
 = - {73\over2520} \pi^6 + (\zeta_3)^2
 = - 26.4048255\ldots,
\label{ESprediction}
\ee
motivating the computation of the four-loop four-gluon scattering
amplitude, and the numerical extraction of $f_0^{(4)}$ from 
it.  The result found~\cite{Neq44},
\be
f_0^{(4)} = -29.335 \pm 0.052,
\label{f04BCDKS}
\ee
and later with much improved precision~\cite{CSV4},
\be
f_0^{(4)} = -29.29473 \pm  0.00005,
\label{f04CSV}
\ee
was consistent, not with \eqn{ESprediction},
but with a version in which the sign of the $(\zeta_3)^2$ term was
flipped,
\be
f_0^{(4)} \Big|_{\rm BES} = 
{1\over4} \hat{\gamma}_K^{(4)} \Big|_{\rm BES}
 = - {73\over2520} \pi^6 - (\zeta_3)^2
 = -29.2947071202\ldots.
\label{BESprediction}
\ee

Remarkably, the latter value was predicted, simultaneously with
ref.~\cite{Neq44}, by Beisert, Eden and Staudacher (BES)~\cite{BES}, 
based on a modified integral equation taking into account a new 
proposal for the dressing factor, with nontrivial effects beginning 
at four loops.  The proposed dressing factor was deduced by using 
its properties at strong-coupling, where it had been known to be 
nontrivial~\cite{DressingStrong}.
Perhaps even more remarkably, the only effect of including the
dressing-factor term on the weak-coupling
expansion of the integral equation, is to make the substitution
$\zeta_{2k+1} \to i \, \zeta_{2k+1}$, which affects only the signs 
of the odd-zeta terms in the perturbative expansion.
At five loops, this sign-flip is 
\bea
f_0^{(5)} \Big|_{\rm ES}
 &=& (887/56700) \pi^8 - 2 \, \zeta_2  \, (\zeta_3)^2
                     - 10 \, \zeta_3 \, \zeta_5
\ =\ 131.21\ldots 
\label{ES5loops}\\
\to\ f_0^{(5)} \Big|_{\rm BES}
 &=& (887/56700) \pi^8 + 2 \, \zeta_2  \, (\zeta_3)^2
                   + 10 \, \zeta_3 \, \zeta_5
\ =\ 165.65\ldots\,,
\label{BES5loops}
\eea
which also agrees with interpolation-based estimates~\cite{Neq44}.

The BES integral equation was solved
numerically~\cite{BESStrongExpNum}, and later expanded analytically
to all orders in the strong-coupling ($1/\sqrt{\lambda}$)
expansion~\cite{BESStrongExpAnal}.  Its strong-coupling behavior is
consistent with the known first three terms in this
expansion~\cite{StrongCouplingLeading,StrongCouplingSubleading,RTT}.
This concordance, plus the agreement with the first four loops at
weak coupling, strongly suggests that the BES equation is an exact
solution for the cusp anomalous dimension, valid for arbitrary $\lambda$.

The next quantity appearing in the planar MSYM gluon scattering amplitudes,
$G_0(\lambda)$, which controls single poles in the argument of the
exponential in \eqn{Mplanarnew}, is not quite as well known.
The first four loop coefficients are known, the fourth 
numerically~\cite{CSVCollinear},
\be
G_0(\lambda) = -\zeta_3 \biggl({\lambda\over8\pi^2}\biggr)^2
+ {2\over3} \bigl( 6 \zeta_5 + 5 \zeta_2\zeta_3 \bigr)
\biggl({\lambda\over8\pi^2}\biggr)^3 
- (77.56 \pm 0.02) \biggl({\lambda\over8\pi^2}\biggr)^4 + \cdots \,, 
\label{G04loops}
\ee
and one coefficient is now known in the strong-coupling
expansion~\cite{AM1}.
A Pad\'e approximant incorporating this data has been
constructed~\cite{CSVCollinear}.   Clearly, it would be of great interest
if an integral equation could be found governing $G_0(\lambda)$
for all values of the coupling.  Finding a cleaner operator interpretation 
for this quantity may be quite useful in this respect.


\section{Gluon scattering at strong coupling}
\label{strongsection}

\begin{figure}
\includegraphics[width=.8\textwidth, bb = -30 570 420 720 ]{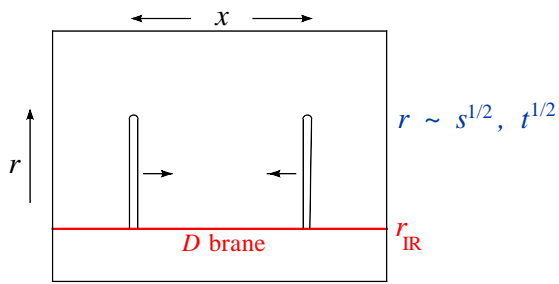}
\caption{Gluon scattering in anti-de Sitter space.  Four-dimensional
  space-time has coordinates $x$. Hard-scattering kinematics force the
strings to stretch a long distance in the radial direction $r$,
from their infrared ``anchor'', a $D$ brane located at $r_{\rm IR}$.}
\label{GluonsAdSFigure}
\end{figure}

Now let us return to the picture of gluon scattering at strong
coupling developed by Alday and Maldacena~\cite{AM1}.
\Fig{GluonsAdSFigure} is another view of the AdS space sketched
in \fig{AdS5sketchFigure}, showing also a pair of incoming open
string states prior to a hard scattering.  The ends of the open
strings are anchored on a $D$ brane, which serves as an infrared
regulator and is located at a small value of the AdS radial variable,
$r_{\rm IR}$.  The short-distance (UV) nature of the hard scattering
forces part of the string to penetrate to large values of 
$r \sim \sqrt{s}, \sqrt{t}$.   Gluons correspond to this part of
the string, and the rest of the string can be thought of as the color
string a gluon has to drag along with it, which is particularly
important at strong coupling.  Because the string has to stretch a long
way, the scattering is semi-classical~\cite{AM1}.

This regime is similar to very high-energy, fixed-angle scattering in 
string theory in flat space-time, which was studied long 
ago~\cite{GrossMende}.
Evaluated on the classical solution, for the case of color-ordered
scattering with gluons 1 and 3 incoming, 2 and 4 outgoing,
the string world-sheet action is imaginary.
The Euclidean action, or area, is real, and is logarithmically
divergent, leading to a large exponential suppression~\cite{AM1},
\be 
{\cal M}_4 \sim \exp[i S_{\rm cl}] \sim \exp[-S^E_{\rm cl}]
 \sim \exp[ - \sqrt{\lambda} \ln^2(r/r_{\rm IR}) ] \,,
\label{expsuppress}
\ee
where $r \sim \sqrt{s}, \sqrt{t}$.
The coupling-constant dependence in \eqn{expsuppress} originates
from the formula for the radius of curvature of AdS, 
$R_{\rm AdS}^2 = \sqrt{\lambda}$, which enters the world-sheet 
action.  From the string point of view, the suppression can
be attributed to a tunnelling suppression factor.
From the point of view of a four-dimensional collider physicist,
it is a typical Sudakov suppression factor~\cite{FormFactors}: 
The probability for a pair of gluons to make it all the way
into and back out of the scattering without radiating at all
is exponentially small --- especially at strong coupling, 
$\lambda\to\infty$ --- with a double log in the exponential.

\begin{figure}
\includegraphics[width=.80\textwidth, bb = 70 500 570 695 ]%
{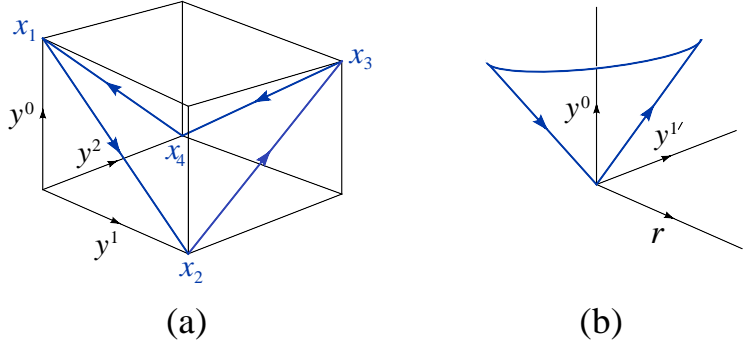}
\caption{(a) Boundary condition at $r=0$ for $gg\to gg$
scattering at $90^\circ$ in the $u$-channel.  
(b) The cusp solution, showing $r$ as a
function of $y^0$ and $y^{1\prime}$.}
\label{YbdyCuspFigure}
\end{figure}

To make contact with the perturbative results, Alday and Maldacena
constructed a dimension\-ally-regularized version of AdS${}_5\times S^5$,
instead of using the $D$ brane location $r_{\rm IR}$ as a regulator.
They also introduced $T$-dual variables $y^\mu$ in place of
the usual four-dimensional coordinates.  The $T$-duality 
transformation is a kind of Fourier transform, so the $y^\mu$
are like momentum variables.  Indeed, the asymptotic boundary value
for the world-sheet, which resides at $r=0$ in the 
dimensionally-regularized setup, is a polygon constructed from 
light-like segments in $y^\mu$, with the corners $y_i^\mu$
satisfying
\be
 y_{i+1}^\mu - y_i^\mu = k_i^\mu \,,
\label{yboundary}
\ee
where $k_i^\mu$ is the momentum of the $i^{\rm th}$ gluon.
From \eqn{yboundary}, we see that the $y_i$ coincide precisely
with the dual variables $x_i$ introduced in 
section~\ref{pseudoconfsection} to discuss dual conformal invariance!
\Fig{YbdyCuspFigure}(a) shows the light-like quadrilateral boundary
satisfying \eqn{yboundary} for the case of $2\to2$ gluon scattering
at $90^\circ$ in the 1-2 plane, with $k_1$ and $k_3$ incoming.
The vertical direction is the (dualized) time direction.

\begin{figure}
\includegraphics[width=.80\textwidth, bb = -40 500 460 715 ]%
{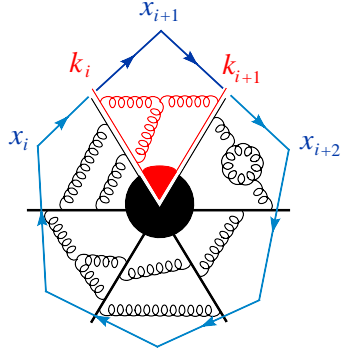}
\caption{Planar Feynman diagrams, ringed by the strong-coupling
boundary condition in dual momentum variables. Each Sudakov wedge
has a single cusp associated to it.}
\label{PlanarExpFigure}
\end{figure}

Near each corner of the polygonal boundary, the solution must
look like a cusp solution, previously constructed by
Kruczenski~\cite{Kruczenski}, in which $r$ behaves like
$r = \sqrt{(2+\e)[(y^0)^2- (y^{1'})^2]} = \sqrt{(2+\e)y^+y^-}$ 
for some spatial coordinate $y^{1'}$, and light-cone coordinates
$y^\pm$.  This hyperboloid is shown in \fig{YbdyCuspFigure}(b).  
The classical action (area) for this solution has a divergence 
regulated by $\e$,
\be 
i S_{\rm cl} = - S^E_{\rm cl} 
\ \to\ - R^2_{\rm AdS} \int_0 { dy^+ dy^- \over (y^+y^-)^{1+\e/2} }
\ \sim\ - {1 \over \e^2} { \sqrt{\lambda} \over 2\pi }
\ \sim\ - { 1\over \e^2 } { \gamma_K(\lambda) \over 2 } \,.
\label{cuspaction}
\ee
The coefficient of the leading divergence is just the 
strong-coupling limit of the cusp anomalous 
dimension~\cite{StrongCouplingLeading},
\be
\gamma_K(\lambda) \sim { \sqrt{\lambda} \over \pi } \,,
\qquad \hbox{as }\lambda\to\infty\,.
\label{cuspstrong}
\ee
\Fig{PlanarExpFigure} illustrates the situation heuristically.
The singular part of the planar amplitude can be broken up
into Sudakov wedges, as in \fig{SoftCollFactPlanarFigure} and 
\eqn{SJhplanar}.  The overlap of soft and collinear divergences
corresponds to regions between two hard lines, {\it e.g.}
$k_i$ and $k_{i+1}$.  Thus each wedge is associated with a 
single divergent cusp~\cite{BuchbinderIR}, of the form shown in 
\fig{YbdyCuspFigure}(b).

The full classical solution, for arbitrary scattering angle, was 
found by Alday and Maldacena~\cite{AM1}.  Its action gives a 
strong-coupling amplitude of the form,
\bea
{\cal M}_4 &=& \exp[-S^E_{\rm cl}]\,,
\nonumber\\
-S^E_{\rm cl} &=& 
- { 1 \over \e^2 } {\sqrt{\lambda} \over \pi}
 \biggl[ \biggl({\mu_{\rm IR}^2 \over -s}\biggr)^{\e/2} 
       + \biggl({\mu_{\rm IR}^2 \over -t}\biggr)^{\e/2} \biggr]
\nonumber\\
&&\null
 -\  { 1 \over \e } {\sqrt{\lambda} \over 2\pi} (1-\ln2)
 \biggl[ \biggl({\mu_{\rm IR}^2 \over -s}\biggr)^{\e/2} 
       + \biggl({\mu_{\rm IR}^2 \over -t}\biggr)^{\e/2} \biggr]
\nonumber\\
&&\null
+ { \lambda \over 8\pi } 
 \biggl[ \ln^2\biggl({s\over t}\biggr) 
       + \tilde{C} \biggr] + \Ord(\e)\,,
\label{AMSE}
\eea
where $\mu_{IR}^2 = 4 \pi e^{-\gamma} \mu^2$.
This expression can be compared with the strong-coupling extrapolation
of the ansatz~(\ref{BDSansatz})~\cite{AM1}.  The $1/\e^2$ poles agree,
using the strong-coupling value for $\gamma_K(\lambda)$ from
\eqn{cuspstrong}.  The $1/\e$ poles give the strong-coupling limit
of the collinear anomalous dimension $G_0(\lambda)$,
\be
G_0(\lambda) \sim \sqrt{\lambda} { (1 - \ln 2) \over 2\pi } \,, 
\qquad \hbox{as }\lambda\to\infty\,.
\label{G0strong}
\ee
The finite part of ${\cal M}_4$ has a dependence on $s$ and $t$ 
which is precisely as predicted by \eqn{BDSfiniteterms}.

\section{Dual variables and Wilson loops at weak coupling}

The dual momentum variables $x_i^\mu$ play a prominent role in 
the strong-coupling
computation of Alday and Maldacena, which is essentially
the same as computing a Wilson loop vacuum expectation value at 
strong coupling.  Inspired by this connection, there has been
a sequence of recent Wilson-loop computations for loops
corresponding to the dual-momentum boundary conditions for
an $n$-point amplitude, namely polygons composed of $n$-light-like
segments, with corners obeying~\eqn{yboundary}.   

The first of these computations was by Drummond, Korchemsky and 
Sokatchev~\cite{DrummondVanishing}, for the one-loop expectation 
value of a quadrilateral ($n=4$) Wilson loop.  Up to constants
of the kinematics, attributable to a different regulator (in the UV)
than the one used for the amplitudes (in the IR),
the expectation value agreed, surprisingly, with the one-loop 
four-gluon amplitude, normalized by the tree amplitude, 
{\it i.e.} \eqn{BDSfiniteterms}.  
Next, Brandhuber, Heslop and Travaglini~\cite{BHT} showed that 
the same statement is actually true for the $n$-gon Wilson loop
for any $n$, compared with the normalized one-loop 
amplitude~\cite{Neq4Oneloop} for the so-called 
maximally-helicity-violating (MHV) configuration of gluon 
helicities (two negative and $(n-2)$ positive).
The Wilson-loop computation knows nothing about the polarizations
of the external gluons.  It is manifestly symmetric under cyclic
permutations and reflections of the polygon.
For $n=4$ and 5, a Ward identity for $\NeqFour$ supersymmetry
shows that all helicity configurations in MSYM are equivalent,
and that the normalized amplitudes have the same manifest 
symmetries as the polygonal Wilson loop~\cite{SWI}.  
However, beyond $n=5$ there are non-MHV configurations which
do not have these symmetries.  How does the Wilson loop know
it is ``supposed to'' match the MHV amplitude alone?

Drummond, Henn, Korchemsky and Sokatchev (DHKS) then 
repeated the Wilson-loop computation in MSYM at two loops, 
first for the $n=4$ case~\cite{DHKSTwoloopBoxWilson} and then\footnote{%
Most of the results reported from this point on appeared after this 
talk was presented, but before the write-up was completed.  
I include them here because of their close connection with the 
contents of the talk.}
for the $n=5$ case~\cite{ConformalWard}.  Again the results 
matched the full two-loop MSYM scattering 
amplitudes~\cite{ABDK,TwoLoopFive}, up to constants of the kinematics.
Furthermore, DHKS first proposed~\cite{DHKSTwoloopBoxWilson}
and then proved~\cite{ConformalWard} an anomalous
dual conformal Ward identity for Wilson loops, in which
the anomaly arises from UV divergences proportional to
$\gamma_K(\lambda)$.   The solution to the Ward identity
is unique for $n=4$ and 5.  Beyond $n=5$, there are multiple
solutions, due to the existence of nontrivial conformally-invariant 
cross ratios.  For example, for $n=6$ the quantity 
$u_1 \equiv x_{13}^2 x_{46}^2/(x_{14}^2 x_{36}^2) =
            s_{12} s_{45}/(s_{123} s_{345})$
is invariant under the inversion~(\ref{inversionsym}),
and there are two other such cross ratios. 
(The appearance of $x_{i,i+1}^2 = k_i^2$ in a cross ratio is 
forbidden by the on-shell constraint $k_i^2=0$.)

DHKS also showed that the amplitude ansatz~(\ref{BDSansatz})
obeys the anomalous dual conformal Ward identity.  
Given that the ansatz was known to be correct for $n=4$ and
5~\cite{ABDK,TwoLoopFive}, and the uniqueness of the Ward identity
solution for these cases, this result could explain why the
amplitude should match the Wilson loop in these cases.
However, it was not clear what should happen for larger $n$.
Indeed, Alday and Maldacena~\cite{AM3} gave an argument, based
on approximating a Euclidean rectangular loop by a zig-zag configuration
composed of many light-like segments, that the 
ansatz~(\ref{BDSansatz}) should fail at strong coupling for 
sufficiently large $n$.   DHKS~\cite{HexagonWilson} 
found that the hexagonal Wilson loop could {\it not} be described at 
two loops by the ansatz~(\ref{BDSansatz}).  This result left open 
the question, however, of whether the ansatz failed to describe MHV 
amplitudes beyond $n=5$, or whether the relation between amplitudes 
and Wilson loops failed beyond two loops (or both).

The high-energy limits of the ansatz~(\ref{BDSansatz}) 
have been examined for consistency with expected Regge 
behavior.  For $n=4$ and 5, the ansatz appears to have consistent
behavior in all such limits~\cite{DrummondVanishing,HE0708,BNST,BLSV}
However, there appears to be a difficulty with the ansatz 
for the six-gluon amplitude starting at two loops~\cite{BLSV}.
Very recently, a computation of the ``parity even'' part of the
six-gluon MHV amplitude~\cite{BDKRSVV} has revealed directly that 
the ansatz~(\ref{BDSansatz}) does fail for $n=6$.  However, a numerical
comparison~\cite{BDKRSVV,WilsonValues} with the corresponding 
hexagonal Wilson loop~\cite{HexagonWilson} shows that the 
MHV-amplitude-Wilson-loop equivalence is still intact at two loops
and $n=6$.  This result means that the scattering amplitude also
obeys the dual conformal Ward identity.  On the other hand,
the solution to the Ward identity is not unique for $n=6$.
Hence some other principle, as yet unidentified, is needed
to explain why MHV amplitudes are equivalent to Wilson loops in MSYM.


\section{Conclusions}

We have seen that gluon scattering amplitudes in planar $\NeqFour$
super-Yang-Mills theory have some remarkable properties.
It appears that the exact forms of the four-gluon and five-gluon 
amplitudes are given by the ansatz~(\ref{BDSansatz}),
which depends only on four different functions of the large-$\Nc$
coupling parameter $\lambda$: $f_0$, $f_1$, $f_2$ and $C$.  
Because an exact solution for one of the four functions --- $f_0$,
the cusp anomalous dimension --- seems to be
in hand~\cite{BES}, perhaps one can say that these cases are 
``$1/4$ solved''.  The fixed dependence of the ansatz~(\ref{BDSansatz}) 
on the scattering angle(s) is apparently related to the uniqueness 
of solutions to a dual conformal Ward identity for $n=4$ and 
5~\cite{DHKSTwoloopBoxWilson,ConformalWard}, and an
equivalence between (MHV) amplitudes and Wilson
lines~\cite{AM1,DrummondVanishing,BHT,DHKSTwoloopBoxWilson,ConformalWard}. 
Although the ansatz~(\ref{BDSansatz}) fails for the MHV six-gluon
amplitude~\cite{BDKRSVV} at two loops, the equivalence remains
valid~\cite{BDKRSVV,WilsonValues}.

There are still many open questions.  Are there simple(r) AdS/operator
interpretations of the other three functions?  Can one find integral
equations for them, based on integrability?  What is the precise
relation between integrability and dual conformal invariance? 
Do non-MHV amplitudes obey any simple patterns, or bear any relation
to Wilson loop expectation values?  From the structure of the
one-loop amplitudes, {\it e.g.} for six gluons~\cite{Neq1Oneloop},
any such relations must be considerably more intricate.
What happens in other conformal theories?
Finally, we can hope that some of these advances may eventually help 
to shed light on scattering amplitudes in other gauge theories, particularly QCD, 
whose understanding --- as exemplified by the other talks at this
symposium --- is vital to the search for new physics at the 
Large Hadron Collider.

\vskip0.4cm
\noindent
{\bf Acknowledgments}\\
\vskip-0.2cm
I am grateful to Stefano Catani, Massimiliano Grazzini and the other
organizers of RADCOR\,2007, for the opportunity to present this talk, 
and for putting together a wonderful conference. 
I also thank Babis Anastasiou, Zvi Bern, Michal Czakon, David Kosower, 
Radu Roiban, Volodya Smirnov, Mark Spradlin, Cristian Vergu and 
Anastasia Volovich for collaboration on some of the topics reviewed here.


\end{document}